\documentclass[12pt]{article}
\usepackage[colorlinks=true,linkcolor=blue, linktoc=page, urlcolor=blue,citecolor=magenta]{hyperref}
\usepackage{amsmath,amscd}
\usepackage{amsfonts}
\usepackage{amssymb}
\usepackage{graphicx,shortvrb}
\usepackage{epsfig}
\usepackage{newlfont}
\usepackage{cite}
\usepackage[utf8]{inputenc}

\def\call{\cal L}

\def\calo{\cal O}

\def\a{\alpha}
\def\b{\beta}
\def\g{\gamma}

\def\h{\eta}
\def\l{\lambda}

\def\f{\phi}

\def\m{\mu}
\def\n{\nu}

\def\O{\Omega}

\def\r{\rho}

\def\s{\sigma}

\def\dnc{\nabla^{(\mathrm{nc})}}
\def\gnc{{\Gamma^{(\mathrm{nc})}}}
\def\nc{{(\mathrm{nc})}}
\def\gr{{(\mathrm{gr})}}

\def\GN{\mathrm{G}_{\mathrm{N}}}

\def\tt{\tau}
\def\lp{\left(}
\def\rp{\right)}
\def\pp{\partial}

\begin{document}

\begin{titlepage}
 \vskip 1.8 cm

\begin{center}{\huge \bf
Newton-Cartan, Galileo-Maxwell and Kaluza-Klein} 
\end{center}
\vskip .3cm
\vskip 1.5cm

\centerline{\large {{\bf Dieter Van den Bleeken and Çağın Yunus}}}

\vskip 1.0cm

\begin{center}

\sl Physics Department, Boğaziçi University\\
 34342 Bebek / Istanbul, TURKEY

\vskip 1.5cm

\texttt{dieter.van@boun.edu.tr, cagin.yunus@boun.edu.tr}

\end{center}

\vskip 1.3cm \centerline{\bf Abstract} \vskip 0.2cm \noindent

We study Kaluza-Klein reduction in Newton-Cartan gravity. In particular we show that dimensional reduction and the nonrelativistic limit commute. The resulting theory contains Galilean electromagnetism and a nonrelativistic scalar. It provides the first example of back-reacted couplings of scalar and vector matter to Newton-Cartan gravity. This back-reaction is interesting as it sources the spatial Ricci curvature, providing an example where nonrelativistic gravity is more than just a Newtonian potential.

\end{titlepage}

\tableofcontents
\section{Introduction}
Newton-Cartan gravity was originally formulated to put Newtonian gravity in a manifestly coordinate invariant form \cite{Cartan:1923zea}. In this formulation it is remarkably similar to general relativity and indeed it was later shown that a careful nonrelativistic limit of the Einstein equations leads directly to Newton-Cartan gravity \cite{Dautcourt:1964,Kunzle:1976,Ehlers:1981}. Recently this classic subject has been of renewed interest due to its appearance in the holographic description of asymptotically non-AdS spaces \cite{Christensen:2013lma,Christensen:2013rfa,Hartong:2014oma,Hartong:2014pma} and applications in condensed matter physics \cite{Son:2013rqa,Geracie:2014nka}.

In its most symmetric form the theory is formulated in terms of a one-form $\tau_\mu$, a metric $h^{\mu\nu}$ and a connection $\nabla_\mu^\nc$, defined on a $(1,d)$ dimensional manifold\footnote{As in this paper various numbers of spatial dimensions are discussed we use the notation $(1,d)$ where $d$ indicates the number of spatial directions and the 1 represents the time direction. In the Lorentzian case the distinction between the two is made by the signature of the metric, in the Galilean case $\tau_\mu$ singles out the time direction.}. These fields are however subject to a number of non-dynamical constraints and hence encode the real degrees of freedom of the theory in a somewhat convoluted way. By solving the constraints one can make the unconstrained fields manifest, be it at the cost of breaking manifest time reparameterization invariance. Although such a gauge-fixing would be a rather arbitrary thing to do in a relativistic theory, it is perfectly natural in a Galilean theory as all observers can agree on an absolute time. One of the unconstrained fields making up Newton-Cartan gravity is the Newtonian gravitational potential $\Phi$, but in addition there is also a spatial metric $h_{ij}$ and a vector field $C_i$. Although most authors are aware of these additional fields they typically tend to do away with them quickly, although there are some exceptions \cite{Dautcourt:1990b, Ehlers97}. They seem to be justified in treating $h_{ij}$ and $C_i$ as less important/physical since at first sight the dynamical equations for these fields appear to be very rigid, at least in the case of the vacuum theory or in the presence of a perfect fluid. The equation for the spatial metric is equivalent to vanishing Ricci curvature, which in 3 spatial dimensions implies that it is locally flat. The vectorfield in turn is forced to be harmonic which with some assumptions on the boundary conditions implies its curvature should vanish. An equal fate would await the Newtonian potential, were it not that its Poisson equation is naturally sourced by mass density.

It is clear then that the difference in standing between the fields is not so much in the equations that govern them but rather in the (non-)appearance of sources. In the case of a perfect fluid this comes about because the pressure, which sources the spatial components of the curvature in the relativistic theory, gets washed away in the nonrelativistic limit and only the mass density remains. But one could wonder if this is a general feature of all types of energy and momentum or if there are exceptions. This question is one of the main motivations for this work.

Probably the main reason why coupling various types of matter and fields to Newton-Cartan gravity has not been investigated thoroughly\footnote{Here we are talking about fully back-reacted couplings, where the effect of these additional fields on the gravitational sector is taken into account. There is a rather extensive literature \cite{Kunzle:1976, Duval:1983pb, Kunzle:1984dt,deMontigny:2003uw, Santos:2004pq, Son:2005rv, Jensen:2014aia, Jensen:2014ama, Geracie:2015dea, Geracie:2015xfa, Fuini:2015yva, Auzzi:2015fgg, Banerjee:2014pya, Banerjee:2014nja, Banerjee:2015tga, Banerjee:2015rca}  on the simpler problem of studying various fields on a Newton-Cartan background, where the effect of these fields on that background is ignored.} (to the best of our knowledge) is that the theory does not have a Lagrangian formulation. This makes it harder to investigate various possibilities consistent with the symmetries of the theory or to decide which couplings are physical and minimal. On the other hand it sounds somewhat strange that this would be of any concern since we know plenty of relativistic examples, can't one simply take the nonrelativistic limit of those? The problem is that this limit is rather subtle and requires a starting ansatz. If one knows the endpoint of the limit it is often easy to show how it arises, but without it it can be hard to come up with a good starting point. This will be illustrated in this work as well, where we reproduce our results from a nonrelativistic limit, but this limit includes a crucial subtlety (a choice of conformal frame) which would have been hard to guess without knowing where we were heading.

\begin{figure}[h]\label{diagram}
\begin{center}
\includegraphics[scale=0.9]{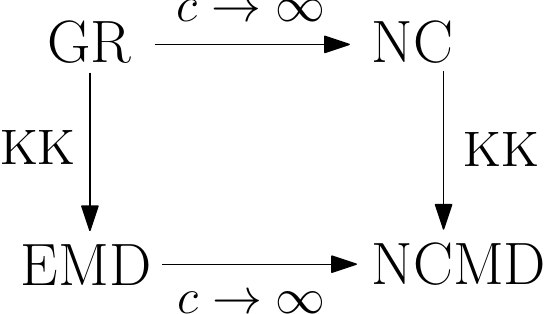}
\end{center}
\caption{This commutative diagram shows the two ways one can go from general relativity (GR) to Newton-Cartan-Maxwell dilaton theory (NCMD). The first option is to pass via Newton-Cartan gravity (NC) by first taking the nonrelativistic, i.e. $c\rightarrow\infty$, limit and then performing Kaluza-Klein reduction (KK). Or one could go via Einstein-Maxwell dilaton theory (EMD) by first performing Kaluza-Klein reduction and then taking the nonrelativistic limit.}
\end{figure}

The way we introduce source terms for the spatial curvature is based on a rather simple idea. A first observation is that the conclusion that Ricci flatness implies flatness only holds in 3 spatial dimensions but not in higher ones. There are plenty of well-known and interesting examples of highly curved Euclidean Ricci flat manifolds. Furthermore we know from the relativistic theory that if such a space has an isometry we can equivalently rewrite it as a solution to a lower dimensional theory through Kaluza-Klein reduction \cite{Bailin:1987jd}. This implies that if this procedure carries through to the nonrelativistic setting there should exist many interesting solutions to Newton-Cartan gravity that have curved spatial directions, even in 3 spatial dimensions. For this to be consistent there needs to a appear a source term for the Ricci curvature, but indeed the reduced theory is expected to have an additional vector and scalar field that could be candidates for such source terms. In the rest of this paper we work out the technical realization of this idea and show that indeed it has all the features one would have expected. In particular we show that it doesn't matter if one first takes a nonrelativistic limit and does dimensional reduction or vice versa, see figure \ref{diagram}.

Apart from the motivation just outlined before this work could be of somewhat more applied interest as well. More elaborate versions of dimensional reduction play an important role in string theory and supergravity in connecting the fundamental UV theory of gravity to our 4 dimensional world. The nonrelativistic limit of these scenarios might lead to an approximation scheme that can introduce some simplification in this rather complicated endeavor. Apart from its phenomenological applications dimensional reduction also represents a powerful unifying and organizing scheme. Instead of needing to study various theories in each dimensions separately it is often sufficient to study some high maximal dimension and by reduction one then obtains a 'web' of lower dimensional theories connected by various dualities. Supersymmetric theories are a good example of this and so it might be that a nonrelativistic understanding of dimensional reduction can help in the search for a supersymmetric extension of Newton-Cartan gravity \cite{Andringa:2013mma,Bergshoeff:2015uaa,Bergshoeff:2015ija} when $d>2$.

Finally we would like to point out that our dimensional reduction of Newton-Cartan gravity is carried through along a spatial direction and in the non-relativistic theory. There is no direct relation with the method of obtaining Newton-Cartan gravity itself via a lightlike reduction of (1,4) dimensional general relativity \cite{Julia:1994bs}.

\paragraph{Overview}

The remainder of this paper is organized as follows. We start in section \ref{NC} by reviewing the basics of Newton-Cartan gravity. As the theory after dimensional reduction will include nonrelativistic electromagnetism we shortly review this theory as well in section \ref{GM} and we put special emphasis on the presence of magnetic charges as the explicit example we study includes those naturally. Section \ref{KK} contains one of the main results of this paper: our reduction ansatz and the equations of motion of the nonrelativistic Kaluza-Klein theory. In section \ref{nonrelsec} we perform the nonrelativistic limit of Einstein-Maxwell dilaton theory, a slight generalization of the Kaluza-Klein reduction of general relativity. We argue that to perform this limit one has to go to a particular conformal frame and present the resulting nonrelativistic equations of motion. In the case one starts with the Kaluza-Klein theory one obtains exactly the same result as the Kaluza-Klein reduction of Newton-Cartan gravity presented in section \ref{KK}, which can be seen as the second main result of the paper. To illustrate that this theory is physically interesting even when the number of spatial dimensions is 3 we present an example in section \ref{example}, namely that of a magnetic monopole, whose presence warps the spatial dimensions into a curved, albeit conformally flat, geometry. 

For the convenience of the reader we have also included two appendices. In the first, appendix \ref{notation}, we summarize some of our conventions and notation. The second, appendix \ref{otherways}, is written for those readers who are more familiar or prefer to work with the formulation of Newton-Cartan gravity that is invariant under the full $(1,d)$ dimensional diffeomorphism group. We present all the theories discussed in the paper; Newton-Cartan, Galileo-Maxwell and Newton-Cartan-Maxwell dilaton, in such a form and explain how the formulation in the main text can be obtained by a particular partial gauge fixing that keeps invariance under time dependent spatial diffeomorphisms intact but breaks time reparametrization invariance. 

\section{Newton-Cartan gravity}\label{NC}
We start by shortly reviewing the structure of Newton-Cartan gravity in $d$ space and 1 time dimensions, i.e. $(1,d)$ dimensions, see e.g. \cite{Havas:1964zza,Kunz:1972,Duval:2009vt,Andringa:2010it} for further details and references. In this Galilean theory there is an absolute time coordinate that all observers can agree on. We will make this explicit by working in a description of the theory where we have gauge-fixed time reparameterizations. We find it natural to proceed this way and feel that it clarifies the role of the different physical fields in the theory, be it at the cost of losing some elegance. For convenience of readers more familiar with the manifestly $(1,d)$ diffeomorphism invariant formulation of the theory we have included appendix \ref{otherways} where we show how the formulation used here is equivalent to these other, more common formulations.

In the formulation here \cite{Dautcourt:1996pm} the theory is manifestly invariant under time-dependent $d$-dimensional\footnote{We use coordinates $x^i$, $i=1\ldots d$ on the spatial manifold. Furthermore we indicate time with $t$ and the time derivative with a dot. See appendix \ref{notation} for more details.} (spatial) diffeomorphisms generated by vectorfields $\xi^i(t,x)$. In addition there is a local $\mathfrak{u}(1)$ gauge symmetry parametrized by $\lambda(t,x)$. Newton-Cartan gravity contains the following three fields
\begin{eqnarray}
\mbox{Newtonian potential:}\qquad&\Phi(t,x)&\qquad\delta \Phi=\call_\xi \Phi-C_i\dot\xi^i-\dot\lambda\label{transfoPhi}\\
\mbox{Coriolis vector:}\qquad& C_i(t,x)&\qquad\delta C_i=\call_\xi C_i+h_{ij}\dot \xi^j+\partial_i\lambda\label{transfoC}\\
\mbox{Spatial metric:}\qquad& h_{ij}(t,x)&\qquad\delta h_{ij}=\call_\xi h_{ij}\label{transfoh}
\end{eqnarray}
Note that contrary to the time reparametrization invariant formulations in terms of constrained fields, the fields listed above are unconstrained.

One sees that when the spatial diffeomorphisms are time-independent, i.e. $\xi^i(x)$, they act as usual purely through the spatial Lie derivative $\call_\xi$. When the spatial diffeomorphism is time dependent however there is an additional action proportional to $\dot \xi^i$, mixing the different fields.  In particular these time dependent diffeomorphisms include local Galilean boosts, $\xi^i=v^it$, and more generally can be interpreted as describing a change to an arbitrary locally non-inertial frame.

The dynamics of these fields is then described by the Newton-Cartan equations of motion:
\begin{eqnarray}
R_{ij}&=&0 \label{eomh}\\
-\nabla^j K_{ji} &=& 2h^{jk}\nabla_{[i}\dot h_{j]k}\label{eomC}\\
-\nabla^iG_i&=&\frac{1}{2}h^{ij}\ddot{h}_{ij}+\frac{1}{4}\dot h_{ij}\dot h^{ij}-\frac{1}{4}K_{ij}K^{ij}+\,4\pi \GN\rho\label{eomPhi}
\end{eqnarray}
Here the covariant derivatives are with respect to the Levi-Civitta connection of the spatial metric $h_{ij}$ and $R_{ij}$ is the Ricci tensor for this connection. Furthermore we introduced the $\mathfrak{u}(1)$-invariant field strengths
\begin{eqnarray}
G_i=-\partial_i\Phi-\dot C_i &\qquad&\delta G_i=\call_\xi G_i+(K_{ij}-\dot h_{ij})\dot\xi^j-h_{ij}\ddot \xi^j \\
K_{ij}=\partial_iC_j-\partial_jC_i &\qquad& \delta K_{ij}=\call_\xi K_{ij}-2h_{k[i}\nabla_{j]}\dot{\xi}^k
\end{eqnarray}
We also included a source term in the form of $\rho$, the mass density, which couples through Newton's constant $\GN$.

Before we continue let us recall that the Newton-Cartan gravity formulated above can be obtained as a $c\rightarrow\infty$ limit of the $(1,d)$ dimensional Einstein equations for gravity coupled to a perfect fluid \cite{Dautcourt:1990, Dautcourt:1996pm, Tichy:2011te}.

Let us also point out to readers not familiar with this nonrelativistic gravity theory that it has the simple solution $h_{ij}=\delta_{ij}$, $C_i=0$ and $\partial_i\partial_i\Phi=4\pi \GN\rho$. This last equation is the well-known Poisson equation for the Newtonian gravitational potential. In case $d=3$ this seems pretty much the most general solution as in that case $R_{ij}=0$ implies that locally there exists a coordinate system such that $h_{ij}=\delta_{ij}$. In the absence of non-trivial boundary conditions \eqref{eomC} then also implies $C_i=0$. This is no longer true in higher dimensions however and we'll show in section \ref{KK} how this can be interpreted as the appearance of source terms in lower dimensions.

\section{Galileo-Maxwell electromagnetism}\label{GM}
It will also be useful to review a few basics about Galilean electromagnetism. This theory was studied as a nonrelativistic limit of standard electromagnetism in \cite{LeBellac1973} where it was formulated as a theory in flat space $h_{ij}=\delta_{ij}, C_i=0$ that is invariant under global Galilean transformations. Recently \cite{Bagchi:2014ysa} investigated the full conformal invariance of this theory. We will be interested in the extension to an arbitrary Newton-Cartan background, which was first performed in \cite{Kunzle:1976}. Some aspects of the inclusion of magnetic charge were discussed in \cite{Strazhev74}.  For a clear discussion of the motivations and applications of nonrelativistic electromagnetism we refer to \cite{LeBellac1973}.

The fundamental fields of Galileo-Maxwell theory are an electric scalar potential $\Psi$ and a magnetic vector potential $A_i$. They have a $\mathfrak{u}(1)$ gauge-transformation and transform as a Galilean 1-form under time-dependent spatial diffeomorphisms:
\begin{eqnarray}
\mbox{Electric potential:}\qquad&\Psi(t,x)&\qquad\delta \Psi=\call_\xi \Phi-A_i\dot\xi^i-\dot\zeta\label{transfopsi}\\
\mbox{Vector potential:}\qquad & A_i(t,x)&\qquad\delta A_i=\call_\xi A_i+\partial_i\zeta\label{transfoA}
\end{eqnarray}
It will be useful to introduce the $\mathfrak{u}(1)$-gauge invariant electric field and magnetic curvature:
\begin{eqnarray}
E_i=-\partial_i\Psi-\dot A_i&\qquad& \delta E_i=\call_\xi E_i+F_{ij}\dot{\xi^j} \label{efield} \\
F_{ij}=\partial_iA_j-\partial_jA_i &\qquad& \delta F_{ij}=\call_\xi F_{ij} \label{bfield}
\end{eqnarray}

The Galilean version of Maxwell's equations on an arbitrary Newton-Cartan background are then
\begin{eqnarray}
\nabla_iE^i=\rho_{\mathrm{(e)}}+\frac{1}{2}K_{ij}F^{ij}&\quad\qquad& \nabla_jF^{ji}=-j_\mathrm{(e)}^i\label{max1}\\
\partial_{[i}F_{jk]}=\rho_{ijk}^\mathrm{(m)}\qquad&\quad\qquad& 2\partial_{[i}E_{j]}=\dot F_{ij}+j_{ij}^{(\mathrm{m})}\label{max2}
\end{eqnarray}
Here the covariant derivatives are with respect to the Levi-Civitta connection of $h_{ij}$. Note however that apart from this 'minimal coupling' of replacing partial derivatives with covariant ones, there is also a coupling to the Coriolis field strength $K_{ij}$ that sources the electric field. We like to point this out since although it is implicitly there in the $(1,d)$ dimensional covariant formulation of \cite{Kunzle:1976} it doesn't get much attention in the literature. We'll see one of its explicit effects in section \ref{example}. One can find the equivalent time-reparametrization invariant form of (\ref{max1},\ref{max2}) in appendix \ref{otherways}. In the formulation here this coupling is required by invariance under the transformations \eqref{transfopsi}.

In the Galileo-Maxwell equations (\ref{max1}, \ref{max2}) we have included both electric and magnetic charge densities and currents as this will also be of some relevance in our example in section \ref{example}. We would like to point out the somewhat unusual conservation equations that follow from these nonrelativistic Maxwell equations:
\begin{equation}
\nabla_i j_{(\mathrm{e})}^i=0\qquad\qquad \dot\rho_{ijk}^{(\mathrm{m})}+\partial_{[i}j^{(\mathrm{m})}_{jk]}=0\label{chcons}
\end{equation}
So although magnetic charge is locally conserved this is not the case for the electric charge, which is only globally conserved, see \cite{LeBellac1973} for further discussion.

Most of the time we will leave the number of spatial dimensions $d$ arbitrary and hence the magnetic charge is not a scalar density nor is its current a vector. When $d=3$ one can easily connect back to the standard formulation via the definitions $\rho^{(\mathrm{m})}=\frac{\sqrt{h}}{2}\epsilon^{ijk}\rho_{ijk}^{(\mathrm{m})}$ and $j^{(m)}_{ij}=\sqrt{h}\epsilon_{ijk}j^{(m)k}$. The conservation equation then takes the familiar form $\dot \rho^{(\mathrm{m})}+\nabla_i j^{(\mathrm{m})i}=0$.

Finally let us point out that in \cite{LeBellac1973}, which is in $d=3$ on flat space, it is stressed that there are two ways to perform the nonrelativistic limit of electromagnetism, leading to what these authors call the electric respectively magnetic limit. If one however includes both magnetic and electric charges one sees that the magnetic and electric limit lead to {\it equivalent} theories that are related by the simple\footnote{Admittedly when expressed in terms of the gauge potentials this field redefinition is non-local, which is a well known feature of electromagnetic duality.}, redefinitions $E_i\rightarrow B_i,\ B_i\rightarrow -E_i$ together with $(\rho_{(\mathrm{e})}, j^i_{(\mathrm{e})})\rightarrow (\rho^{(m)},j^{(\mathrm{m})i}),\ (\rho^{(\mathrm{m})}, j^{(\mathrm{m})i})\rightarrow (-\rho_{(e)},-j_{(\mathrm{e})}^i)$, where $B_i=-\epsilon_{ijk}F_{jk}$. So we prefer to speak of the electric or magnetic {\it formulation} of the nonrelativistic limit rather than use the terminology electric vs magnetic limit. In this language the equations (\ref{max1}, \ref{max2}) together with the definitions (\ref{efield},\ref{bfield}) constitute the magnetic formulation of Galilean electromagnetism.

\section{Kaluza-Klein reduction}\label{KK}
After having introduced some of the main nonrelativistic classical field theories we are now ready to come to the main point of the paper. In this section we will start with Newton-Cartan gravity, as described in section \ref{NC}, in $(1,d+1)$ dimensions. We'll consider field configurations that are independent of one spatial direction and show that under this assumption the $(1,d+1)$ dimensional theory is equivalent to $(1,d)$ dimensional Newton-Cartan gravity coupled to a scalar and electromagnetic field. This is of course nothing but the nonrelativistic version of the classic Kaluza-Klein procedure, see \cite{Bailin:1987jd} for a review.

In this section we will denote the $(1,d+1)$ dimensional fields with a hat, and the $(1,d)$ dimensional ones without. We will split the spatial coordinates as $(x^i,y)$, $i=1,\ldots,d$ where we assume all fields to be independent of $y$. Our reduction ansatz is then
\begin{eqnarray}
\hat\Phi(t,x,y)&=&\Phi(t,x)-\frac{1}{2}\Omega^2(t,x)\Psi(t,x)^2\\
\hat C_y(t,x,y)&=&-\Omega^2(t,x)\Psi(t,x)\\
\hat C_i(t,x,y)&=&C_i(t,x)-\Omega^2(t,x)\Psi(t,x)A_i(t,x)\\
\hat{h}_{yy}(t,x,y)&=&\Omega^2(t,x)\label{ans1}\\
\hat{h}_{iy}(t,x,y)&=&\Omega^2(t,x)A_i(t,x)\\
\hat{h}_{ij}(t,x,y)&=&\Omega^a(t,x)h_{ij}(t,x)+\Omega^2(t,x)A_i(t,x)A_j(t,x)\label{ans2}
\end{eqnarray}
Naively one could assume the parameter $a$ in the equation above to be any constant but we will see immediately that the symmetries fix it to be zero. Indeed not only this constant but the complete structure of the ansatz is fixed by demanding that the lower dimensional fields transform properly under $y$-independent gauge transformations. For gauge parameters $\lambda(t,x)$ and $\hat\xi^{\hat i}(t,x)=(\xi^i(t,x),\zeta(t,x))$ the decomposition of the fields presented above implies the presence of the following lower dimensional fields and transformation laws
\begin{eqnarray}
\mbox{Newtonian potential:}\qquad&\Phi(t,x)&\qquad\delta \Phi=\call_\xi \Phi-C_i\dot\xi^i-\dot\lambda\\
\mbox{Coriolis vector:}\qquad& C_i(t,x)&\qquad\delta C_i=\call_\xi C_i+\Omega^{a}h_{ij}\dot \xi^j+\partial_i\lambda\\
\mbox{Spatial metric:}\qquad& h_{ij}(t,x)&\qquad\delta h_{ij}=\call_\xi h_{ij}\\
\mbox{KK potential:}\qquad&\Psi(t,x)&\qquad\delta \Psi=\call_\xi\Psi-A_i\dot\xi^i-\dot\zeta\\
\mbox{KK vector:}\qquad&A_i(t,x)&\qquad\delta A_i=\call_\xi A_i+\partial_i\zeta\\
\mbox{Radion:}\qquad&\Omega(t,x)&\qquad\delta \Omega=\call_\xi\Omega
\end{eqnarray}
One observes now that the lower dimensional fields $(\Phi, C_i, h_{ij})$ transform properly as the Newton-Cartan fields of section \ref{NC} only\footnote{Note that redefining $\tilde C_i=\Omega^a C_i$ would cure the $\xi$ transformation of $h_{ij}$ but would in turn give the wrong $\lambda$ transformation.} if we take
\begin{equation}
a=0\,.\label{ais0}
\end{equation}
Once the ansatz has been fixed it is a matter of algebra to recast the $(1,d+1)$ dimensional equations of motion as an equivalent set of $d$-dimensional equations of motion. The result is
\begin{eqnarray}
\nabla_i\partial^i\Omega&=&\frac{1}{4}\Omega^{3}F_{ij}F^{ij}\nonumber\\
\nabla_i\left(\Omega^3E^i\right)&=&\frac{1}{2}\Omega^3K_{ij}F^{ij}\nonumber\\
\nabla_i\left(\Omega^3F^{ij}\right)&=&0\label{KKeqs}\\
\Omega R_{ij}&=&\nabla_i\partial_j\Omega+\frac{1}{2}\Omega^3F_{i}{}^kF_{jk}\nonumber\\
\nabla_j\left(\Omega K^j{}_{i}\right)&=& -\O \,\partial_i (h^{jk}\dot h_{jk}) +\nabla^j (\O \dot h_{ij}) -2 \partial_i \dot \Omega+\O^3 F_{ij}E^j\nonumber \\ -\nabla_i\left(\Omega G^i\right)&=&4\pi \GN\rho+\frac{1}{4}\Omega\left(\dot h^{ij}\dot h_{ij}-K_{ij}K^{ij}+2h^{ij}\ddot h_{ij}\right)+\frac{1}{2}\Omega^3E_iE^i+\ddot{\Omega}\nonumber
\end{eqnarray}
Let us point out that to obtain the last equation we also used that Newton's constant and the mass density get rescaled in the reduction: $\hat{\mathrm{G}}_\mathrm{N}\hat{\rho}=\Omega^{-1}\GN\rho$, with $\hat{\mathrm{G}}_\mathrm{N}=2\pi R\, \GN$, where $R$ is the radius of the internal dimension at spatial infinity.

The equations above present a generalization of both Newton-Cartan gravity and Galilean electromagnetism. The first three equations represent a Galilean scalar and electromagnetic field on an arbitrary Newton-Cartan background, as reviewed in section \ref{GM}. More interestingly we see that these fields also back-react and appear as non-trivial sources in the right hand sides of the last three equations describing  the Newton-Cartan fields. A fully $(1,d)$ diffeomorphism invariant version of these equations is presented in appendix \ref{otherways} for completion.

We will present an interesting non-trivial solution to the above theory in section \ref{example} but first we show how it can also be obtained as a nonrelativistic limit of ordinary Kaluza-Klein theory.

\section{Non-relativistic limit}\label{nonrelsec}
We start with a generic, relativistic Einstein-Maxwell dilaton theory in $(1,d)$ space-time dimensions:
\begin{equation}
L = \sqrt{-\tilde g} \lp \tilde R^{(\mathrm{gr})}-\tilde g^{\mu\nu}\frac{1}{2}\pp_\m \f \pp_\n \f - \frac{1}{4}e^{-2k\f}\tilde H^2\rp \label{EMD}
\end{equation}
Here $\tilde R^{(\mathrm{gr})}$ is the standard (general relativistic) Ricci scalar computed for the Lorentzian metric $\tilde g_{\mu\nu}$ through its Levi-Cevitta connection and $\tilde H^2=\tilde g^{\mu\nu}\tilde g^{\rho\sigma}H_{\mu\rho}H_{\nu\sigma}$ with $H_{\mu\nu}=\partial_\mu B_\nu-\partial_\nu B_\mu$. The theories are parameterized by a constant $k$ that can be freely chosen. The theory is equivalent to the Kaluza-Klein reduction of $(1,d+1)$ dimensional Einstein gravity if one chooses 
\begin{equation}
k=\sqrt{\frac{d}{2(d-1)}}\label{kkk}
\end{equation}

As we will argue below the nonrelativistic limit we perform is restricted to a particular conformal frame. We start by performing a generic Weyl rescaling of the metric using the scalar $\f$, parameterized by a constant $l$:
\begin{equation}
\tilde g_{\mu\nu} = e^{2l \f} g_{\mu\nu} \label{Weylredef}
\end{equation} 
Written in terms of the new metric the equations of motion following from \eqref{EMD} are equivalent to
\begin{eqnarray}
R^\gr{}_{\m \n } &=& \lp \frac{1}{2} + (1-d)l^2 \rp \pp_\m \f \pp_\n \f + l(d-1) \nabla_\m \partial_\n \f \nonumber \\
&&+\frac{1}{2} g_{\m \n} \lp -k l + \frac{1}{2(1-d)}  \rp e^{-2(k+l)\f}H^2   \label{relric}\\ 
&&+ \frac{1}{2} e^{-2(k+l)\f} g^{\rho\sigma}H_{\m \r}H_{\n\sigma} \nonumber \\
g^{\m\n}\nabla_\m\left( e^{-\lp 2k+(3-d)l \rp\f}H_{\n \rho}\right) &=&  0\\
g^{\mu\nu}\nabla_\mu\partial_\nu\f &=& (1-d)l \,g^{\m\n}\partial_\mu\f\partial_\nu \f - \frac{k}{2}e^{-2(k+l)\f}H^2\label{relphi} 
\end{eqnarray}
Here all covariant derivatives and the Ricci tensor $R^\gr{}_{\m \n }$ are with respect to the Levi-Cevitta connection of $g_{\m\n}$, as is $H^2=g^{\mu\nu}g^{\rho\sigma}H_{\mu\rho}H_{\nu\sigma}$.

We will now perform the nonrelativistic limit following Dautcourt \cite{Dautcourt:1964,Kunzle:1976, Dautcourt:1990, Dautcourt:1996pm,Tichy:2011te}, with the important subtlety that we apply his expansion in inverse powers of $c$ to the Lorentzian metric $g_{\mu\nu}$ which for $l\neq 0$ is not the Einstein-frame metric $\tilde g_{\m\nu}$ that appears in  the standard form of the Lagrangian \eqref{EMD}. We write $\eta=c^{-2}$ and make the following expansion ansatz for the various fields\footnote{For the strict $c\rightarrow\infty$ limit it is sufficient to make an expansion in even powers of $c$. If one wants to compute the large $c$ corrections to this limit one can take into account the odd powers as well. See \cite{Dautcourt:1990,Dautcourt:1996pm} for more details. Let us also point out that we put the first subleading terms of $\phi$ and $B_\mu$ to zero by hand. This will turn out to be a consistent choice, although one is not forced to do so.}:
\begin{eqnarray}
g_{\m \n} &=& \a_{\m \n} \h^{-1} + \g_{\m \n} + \calo(\eta)\label{exp1} \\
g^{\m \n} &=& h^{\m \n} + \b^{\m \n}\h  + \calo(\eta^2)\\
\phi&=& \varphi+\calo(\eta^2)\\
B_\mu&=& A_\mu+\calo(\eta^2)\label{exp2}
\end{eqnarray}
It is assumed a priory that $h^{\m \n}$ is positive definite with a single zero eigenvalue. The fact that $g^{\mu\nu}$ is the inverse of $g_{\mu\nu}$ imposes then the constraints
\begin{equation}
\a_{\m \n} = -\tau_\m \tau_\n \qquad \qquad h^{\mu\nu}\tau_\nu=0 \qquad h^{\m \n} \g_{\n \r}-
\b^{\m \n} \tau_\nu\tau_\rho = \delta^\m{}_\r \label{limconstraints}
\end{equation}
From this ansatz one can compute the curvatures and one finds that
\begin{eqnarray}
R^\gr{}_{\m \n} &=&R^{(-2)}_{\m \n} \h^{-2}  + R^{(-1)}_{\m \n}\h^{-1} + R_{\m \n}^{(0)}+\calo(\eta) \\
H_{\m\n}&=&F_{\mu\nu}+\calo(\eta)
\end{eqnarray}
Note that here $R_{\mu\nu}^{(\cdot)}$ are {\it not} the Ricci tensors of something, they are by definition the coefficients appearing in the expansion above. The same is true for $F_{\m\nu}$ but it follows from our expansion ansatz that $F_{\mu\nu}=\partial_\mu A_\n-\partial_\nu A_\m$.

We can now study the equations of motion (\ref{relric}-\ref{relphi}) in this expansion.  To start one observes that the right hand side of \eqref{relric} has no term of order $\eta^{-2}$ so it follows that $R^{(-2)}_{\m \n}$ must vanish. A short computation shows that this condition is equivalent to $h^{\m\n} h^{\r \s} \partial_{[\m}\tt_{\r]} \partial_{[\nu}\tt_{,\s]}=0$, which is solved by $\tau_\mu=f \partial_\mu t$. If one demands, like in \cite{Dautcourt:1964, Kunzle:1976}, that the Levi-Civatta connection of $g_{\mu\nu}$ should remain non-singular in the $\eta\rightarrow 0$ limit, then one is forced to choose $f$ to be a constant\footnote{It is an interesting possibility to investigate the nonrelativistic limit without this constraint, something we plan to return to in the future.}, which in turn can be put to 1 by redefining $t$. This then has two consequences.  

Firstly one can check that it makes $R^{(-1)}_{\mu\nu}$ vanish. Comparing with the right hand side of $\eqref{relric}$, of which the term proportional to $g_{\m\nu}$ is of order $\eta^{-1}$, we learn that the limit is only consistent if we choose
\begin{equation}
l=\frac{1}{2k(1-d)}\label{lchoice}
\end{equation} 
Via \eqref{Weylredef} this has the interpretation of fixing a particular conformal frame. Note that the above choice doesn't work for the case $k=0$, which coincides with the rather interesting case of Einstein-Maxwell theory.

Secondly the choice $f=1$ implies we can choose coordinates such that $t=x^0$ and hence $\tau_\mu=\delta^0_\mu$. One can then solve the constraints \eqref{limconstraints} by 
\begin{equation}
\beta^{00}=1\,,\quad \beta^{0i}=-h^{ij}\gamma_{j0}\,,\quad \gamma_{ij}=h_{ij}\,.
\end{equation}

These considerations take care of all inverse powers of $\eta$ and so the leading terms in the equations of motion are of order $\calo(1)$. One can now work out these zeroth order equations of motion in terms of the coefficients in the expansions (\ref{exp1}-\ref{exp2}).  To compare with our earlier discussions on Newton-Cartan gravity we rename $A_0=-\Psi$, $\gamma_{i0}=C_i$, $\gamma_{00}=-2\Phi$ and $e^{-\frac{\varphi}{2k}}=\Omega$. A straightforward calculation then shows that these equations are equivalent to 
\begin{eqnarray}
\nabla_i\partial^i\Omega&=&\frac{1}{4}\Omega^{q+3}F_{ij}F^{ij}\nonumber\\
\nabla_i\left(\Omega^{2q+3}E^i\right)&=&\frac{1}{2}\Omega^{2q+3}K_{ij}F^{ij}\nonumber\\
\nabla_i\left(\Omega^{2q+3}F^{ij}\right)&=&0\label{nonreleqs}\\
\Omega R_{ij}&=&\nabla_i\partial_j\Omega+q\Omega^{-1}\partial_i{\Omega}\partial_j\Omega+\frac{1}{2}\Omega^{q+3}F_{i}{}^kF_{jk}\nonumber\\
\nabla_j\left(\Omega K^j{}_{i}\right)&=& -\O \,\partial_i (h^{jk}\dot h_{jk}) +\nabla^j (\O \dot h_{ij}) -2 \partial_i \dot \Omega-2q\Omega^{-1}\dot{\Omega}\partial_i\Omega+\O^{q+3} F_{ij}E^j\nonumber \\ -\nabla_i\left(\Omega G^i\right)&=&\frac{1}{4}\Omega\left(\dot h^{ij}\dot h_{ij}-K_{ij}K^{ij}+2h^{ij}\ddot h_{ij}\right)+\frac{1}{2}\Omega^{q+3}E_iE^i+\ddot{\Omega}+q\Omega^{-1}\dot{\Omega}^2\nonumber
\end{eqnarray}
where
\begin{equation}
q=2k^2-\frac{d}{d-1}
\end{equation}
For those readers preferring the more elegant time-reparametrization invariant form of the above equations we have provided this in appendix \ref{otherways}. 

Note that the case of Kaluza-Klein reduction corresponds to \eqref{kkk} which implies $q=0$ and then the above equations coincide exactly with those presented in \eqref{KKeqs}.  This thus explicitly establishes the commutative diagram of figure \ref{diagram}.

In hindsight the fact that the reduction and the nonrelativistic limit commute clarifies a few things. Starting with a relativistic Kaluza-Klein ansatz and expanding it in powers of $\eta$ leads to the non-relativistic ansatz (\ref{ans1}-\ref{ans2}), if one chooses the correct conformal frame. In the limiting procedure the conformal frame gets fixed by the demand of a non-singular limit through \eqref{lchoice}. This is equivalent to the condition \eqref{ais0} in the nonrelativistic reduction which there followed from compatibility with the symmetries of the non-relativistic theory.

Finally we would like to point out that our application of Dautcourt's limiting procedure vindicates its usefulness. It is a nonrelativistic limit that does not a priory start with an expansion around Minkowski space, in contrast to the more standard post-Newtonian expansion \cite{Poisson:2014}. When one considers only perfect fluid matter, as Dautcourt did, the resulting nonrelativistic equations of motion force the zeroth order spatial metric to be flat. This then implies that the leading part of the metric is Minkowski after all, and the nonrelativistic limit then coincides with the post-Newtonian expansion. This continues to hold for subleading orders \cite{Dautcourt:1996pm}. Here, where we also consider scalar and vector matter, there are nonrelativistic solutions with a non-flat spatial metric, which implies the zeroth order metric is {\it not} Minkowski space and hence describes a limiting sector of Einstein-Maxwell-dilaton theory that is {\it not} captured by the post-Newtonian expansion. Indeed, translated back to the original relativistic Einstein-frame metric, the zeroth order nonrelativistic solution corresponds to
\begin{equation}
d\tilde s^2=\Omega\left(-(c^2+2\Phi)dt^2+C_idx^idt+h_{ij}dx^ix^j\right)+\calo(c^{-2})\label{dexp}
\end{equation}
while the post-Newtonian expansion \cite{Poisson:2014} assumes metrics of the form
\begin{equation}
d\tilde s^2=-(c^2+2\Phi)dt^2+dx^idx^i+\calo(c^{-2})\label{pnewt}
\end{equation}
So nonrelativistic solutions with non-trivial $\Omega$, $C_i$ or $h_{ij}$ fall outside the standard post-Newtonian expansion, see the next section for an explicit example. 

\section{Example}\label{example}
We end our paper with an explicit example, to show that the new nonrelativistic theory we presented is not a vacuous construction. We provide a solution of the reduced theory with a non-trivial source for the spatial curvature and discuss its physical interpretation as a magnetic monopole, with possible time dependent magnetic charge.

Both for simplicity and physical relevance we restrict ourselves here to $d=3$. In this case the theory \eqref{KKeqs} is nothing but the dimensional reduction of pure Newton-Cartan theory in $(1,4)$ dimensions so we know that the solution should originate from a Ricci flat 4d Euclidean metric with an isometry\footnote{In \cite{Dunajski:2015lxa} such Ricci flat metrics have been studied in the non-relativistic setting, but in a slightly different context. There they are form all of a 4d space-time, interpolating between relativistic and non-relativistic. Here those manifolds form only the spatial part and directly of a non-relativistic 5d geometry.}. These type of metrics are well-studied and have been fully classified in \cite{Boyer:1982mm,Gegenberg84}. They come in two classes, the Killing vector can be of so called rotational or translational type. Here we will restrict ourselves to the second case which correspond to metrics also known as the Gibbons-Hawking metrics \cite{Gibbons:1979zt}:
\begin{equation}
d\hat{s}^2_{(4)}=Vdx^idx^i+V^{-1}(dy+\omega_idx^i)^2
\end{equation}
This metric is Ricci flat when $\partial_i\omega_j-\partial_j\omega_i=\epsilon_{ijk}\partial_kV$, which in particular implies that $V$ is harmonic: $\partial_i\partial_iV=0$.

We can now simply read of fthe lower dimensional fields from the reduction ansatz (\ref{ans1}-\ref{ans2}):
\begin{equation}
h_{ij}=V\delta_{ij}\,,\qquad \Omega=V^{-1/2}\,,\qquad A_i=\omega_i\quad\Rightarrow\quad F_{ij}=\epsilon_{ijk}\partial_k V
\end{equation}
From this one can compute that
\begin{eqnarray*}
\nabla_i\partial_jV^{-1/2}&=&\frac{1}{4}V^{-5/2}\left(5\partial_iV\partial_jV-\delta_{ij}\partial_kV\partial_kV-2V\partial_i\partial_jV\right)\\
F_{i}{}^kF_{jk}&=&V^{-1}\left(\delta_{ij}\partial_kV\partial_kV-\partial_iV\partial_jV\right)\\
R_{ij}&=&\frac{1}{4}V^{-2}\left(3\partial_iV\partial_j V+\delta_{ij}\partial_kV\partial_k V-2V\partial_i\partial_j V-2V\delta_{ij}\partial_k\partial_kV\right)
\end{eqnarray*}
One then sees that the Ricci equation, the equation for $\Omega$ and that for $F_{ij}$ in \eqref{KKeqs} all become equivalent to $\partial_i\partial_iV=0$ as expected. 

The remaining equations for $K_{ij}$, $E_i$ and $G_i$ in \eqref{KKeqs} can then be written as:
\begin{eqnarray}
\partial_iE_i&=&(E_i+K_i)V^{-1}\partial_iV\\
\epsilon_{ijk}\partial_jK_k&=&-\partial_i\dot V+\epsilon_{ijk}V^{-1}\partial_jV(E_k+K_k)\\
\partial_iG_i&=&\frac{1}{2}V^{-1}\left(K_iK^i-E_iE^i\right)
\end{eqnarray}
where for convenience we defined $K_i=\frac{1}{2}\epsilon_{ijk}K_{jk}$ and took $\rho=0$. These equations can be solved by taking $K_i=-E_i$, $G_i=0$ and
\begin{equation}
\partial_iE_i=0\qquad \epsilon_{ijk}\partial_{j}E_k=-\partial_i\dot V\label{dotsource}
\end{equation}

As a last step let us make everything completely explicit in the simplest example, that of Euclidean Taub-NUT\footnote{We assume $y$ to be periodic, $y\simeq y+2\pi R$, and the quantized magnetic charge is $p^{(\mathrm{quant})}=\frac{p}{R}$.} :
\begin{equation}
V=1+\frac{p}{r}\,,\qquad \qquad (r^2=x^ix^i)\,.
\end{equation}
Note that $p$ is a physical parameter of the solution that can {\it not} be absorbed in a coordinate transformation. In the lower dimensional theory this solution corresponds to a magnetic monopole, with scalar hair and a warped spatial metric:
\begin{equation}
B_i=\frac{1}{2}\epsilon_{ijk}F_{jk}=-p\frac{x_i}{r^3}\,,\qquad h_{ij}=\left(1+\frac{p}{r}\right)\delta_{ij}\,\qquad \Omega=\sqrt{\frac{r}{r+p}}\,. 
\end{equation}

In case $p$ is a constant one can take $E_i=-K_i=0$. We can make the solution more interesting however by choosing the charge to be a function of time: $p(t)$. In that case $\partial_i\dot V=-\frac{\dot p}{r^3}r_i$ and one then finds the following solutions for $E_i=-K_i$:
\begin{equation}
E_i=\epsilon_{ijk}\frac{\dot p\, n_jx_k}{r(r-x_ln_l)}
\end{equation}
where $n_i$ as an arbitrary unit vector: $n_in_i=1$.

Note that the electric field blows up along the semi-infinite line determined by $x_i=rn_i$. This has the physical interpretation as the presence of a magnetic current that sources the electric field and is required to be there by the magnetic charge conservation equation \eqref{chcons}.

Finally we point out that this example provides a metric that is a zeroth order solution to the Einstein-Maxwell-dilaton theory \eqref{EMD} in a nonrelativistic expansion via \eqref{dexp}:
\begin{equation}
d\tilde s^2=-c^2V^{-\frac{1}{2}}dt^2+V^{\frac{1}{2}}dx^idx^i+\calo(c^{-2})
\end{equation}
Note that this type of solutions is not captured by the post-Newtonian expansion \eqref{pnewt}.

\section*{Acknowledgements}
We thank Eric Bergshoeff and Jan Rosseel for valuable discussions and communication. DVdB is partially supported by TUBITAK grant 113F164 and by the Bo\u{g}azi\c{c}i University Research Fund under grant number 13B03SUP7. 
\appendix
\section{Notation and conventions}\label{notation}
In this appendix we collect a few notations and conventions.

\paragraph{Space and time}
We will denote with $(1,d)$ the presence of $d$ spatial and 1 time direction for which we use coordinates $x^\mu=(t,x^i)$, i.e. $\mu=0,\ldots d$, $i=1,\ldots d$ and $x^0=t$. In the part of the paper where we perform the dimensional reduction we will further split the spatial directions, in $(1,d+1)$ dimensions we write $\hat{x}^{\hat{i}}=(x^i,y)$ where $i=1,\ldots,d$ and  $\hat{i}=1,\ldots,d+1$.

In most of the paper we work in a notation that is only covariant under $d$ dimensional time dependent coordinate transformations, instead of under full $(1,d)$ dimensional coordinate transformations. As this is a language not commonly used in the literature let us point out a few subtleties.
\paragraph{Time derivative}
We will denote the time derivative often with a dot. Our convention is that for a tensor
\begin{equation}
\dot{T}_{i_1\ldots i_m}^{j_1\ldots j_n}=\partial_t\left( T_{i_1\ldots i_m}^{j_1\ldots j_n}\right)  
\end{equation}
As the metric can be time dependent one {\it cannot} use the metric to raise or lower indices on a tensor with a dot. For example: $\dot h^{ij}$ means the time derivative of the inverse metric. This is not equal to the tensor obtained by raising the indices on the time derivative of the metric:
\begin{equation}
\dot h^{ij}=-h^{ik}h^{jl}\dot h_{kl}
\end{equation}

\paragraph{Lie derivative}
We will denote the standard $(1,d)$-dimensional Lie-derivative of a $(1,d)$-dimensional tensor by a $(1,d)$-dimensional vectorfield $\xi^\mu$ as
\begin{equation}
\pounds_\xi T_{\mu_1\ldots\mu_m}^{\nu_1\ldots\nu_n}=\xi^\lambda\partial_\lambda T_{\mu_1\ldots\mu_m}^{\nu_1\ldots\nu_n}+T_{\lambda\mu_2\ldots\mu_m}^{\nu_1\ldots\nu_n}\partial_{\mu_1}\xi^\lambda+\ldots\ -T_{\mu_1\ldots\mu_m}^{\lambda\nu_2\ldots\nu_n}\partial_\lambda\xi^{\nu_1}-\ldots
\end{equation}
Most of the time we will however work with the $d$-dimensional, purely spatial, Lie-derivative. We define it with respect to a possibly time dependent $d$-dimensional vector field $\xi^i$ acting on a possibly time dependent  $d$-dimensional tensor as
\begin{equation}
\call_\xi T_{i_1\ldots i_m}^{j_1\ldots j_n}=\xi^k\partial_k T_{i_1\ldots i_m}^{j_1\ldots j_n}+T_{ki_2\ldots i_m}^{j_1\ldots j_n}\partial_{i_1}\xi^k+\ldots\ -T_{i_1\ldots i_m}^{kj_2\ldots j_n}\partial_k\xi^{j_1}-\ldots
\end{equation}
In case we choose $\xi^\mu=(0,\xi^i)$ the two definitions are simply related, for example in the case of a vector $V^\mu=(W,V^i)$ or a one-form $\omega_\mu=(\sigma,\omega_i)$:
\begin{eqnarray}
\pounds_\xi W&=&\call_\xi W\\
\pounds_{\xi} V^i&=&\call_\xi V^i-W\dot{\xi}^i\\
\pounds_\xi \sigma&=&\call_\xi \sigma+\omega_i\dot{\xi}^i\\
\pounds_\xi \omega_i&=&\call_\xi \omega_i
\end{eqnarray}

\section{Time reparametrization invariant formulations}\label{otherways}
In this appendix we give a manifestly $(1,d)$ dimensional diffeomorphism invariant form of all the equations of motion presented in the main text. We also explain how the formulations there can be obtained from those in this appendix by a partial gauge fixing that leaves time dependent spatial diffeomorphisms manifest. 

\subsection{Metric formulation of Newton-Cartan gravity}\label{mform}
We loosely follow  \cite{Havas:1964zza,Kunz:1972,Andringa:2010it} and refer to these texts for further details and references.
\paragraph{Setup} In the standard 'metric' formulation of Newton-Cartan gravity one starts with the following fields:
\begin{equation}
\tau_\mu\,,\qquad h^{\mu\nu}=h^{\nu\mu}\,,\qquad \gnc_{\mu\nu}^\rho=\gnc_{\nu\mu}^\rho\,. \label{ncfields}
\end{equation}
They transform as two tensors and a connection under $(1,d)$ dimensional coordinate transformations $\delta_\xi x^\mu=-\xi^\mu$:
\begin{equation}
\delta_\xi \tau_\mu=\pounds_\xi \tau_{\mu}\,,\qquad \delta_\xi h^{\mu\nu}=\pounds_\xi h^{\mu\nu}\,,\qquad \delta_\xi \gnc_{\mu\nu}^\rho=\pounds_\xi\gnc_{\mu\nu}^\rho+\partial_\mu\partial_\nu\xi^\rho
\end{equation}
These fields are subject to the following covariant constraints:
\begin{equation}
\tau_\mu h^{\mu\nu}=0\qquad \dnc_\mu \tau_\nu=\dnc_\mu h^{\nu\rho}=0\label{mformconstraints}
\end{equation}
Note that by the symmetry of the connection these constraints furthermore imply that $\partial_{[\mu}\tau_{\nu]}=0$.

The equations of motion are
\begin{equation}
R_{\mu\nu}^\nc=4\pi \GN\,\rho\,\tau_\mu\tau_\nu\qquad h^{\lambda[\mu}R^{\nc\,\nu]}{}_{(\rho\sigma)\lambda}=0\label{ncmetriceom}
\end{equation}
where the above are the Ricci and Riemann tensor for the connection $\nabla^{(\mathrm{nc})}_\mu$, $\rho$ is the mass density and $\mathrm{G}_{\mathrm{N}}$ is Newton's constant. 

\paragraph{Exhibiting the degrees of freedom}
The constraints in \eqref{mformconstraints} involving the covariant derivative reduce the degrees of freedom contained in the connection, but do so differently than in the familiar case of general relativity.  In the relativistic case, where $g^{\m\nu}$ is invertible, the constraint of compatibility between the metric and the connection leads to a unique expression of the connection in terms of the metric and hence the connection contains no additional degrees of freedom. Here the situation is different, but one can continue analogously by solving for the dependence of the connection on $\tau_\mu$ and $h^{\mu\nu}$ explicitly. To do so one introduces two new tensor fields: $\tau^\mu$ and $h_{\mu\nu}=h_{\nu\mu}$:
\begin{equation}
\delta_\xi \tau^\mu=\pounds_\xi\tau^\mu\qquad \delta_\xi h_{\mu\nu}=\pounds_\xi h_{\mu\nu}
\end{equation} These are defined so as to satisfy their own set of constraints:
\begin{equation}
h^{\mu\nu}h_{\nu\rho}+\tau^\mu\tau_\rho=\delta_\rho^\mu\,\qquad \tau^\rho\tau^\sigma h_{\rho\sigma}=0\label{newfieldconstr}
\end{equation}
Note that these imply $\tau_\mu\tau^\mu=1$ and $h_{\mu\nu}\tau^\nu=0$. It is important to realize that these constraints don't fix the new fields completely in terms of $h^{\mu\n}$ and $\tau_\mu$. This follows from the observation that the constraints \eqref{newfieldconstr} are invariant under the following local symmetry \cite{Duval:1983pb}:
\begin{equation}
\delta_\chi h^{\mu\nu}=\delta_\chi\tau_\mu=0\,,\qquad \delta_\chi\tau^\mu=-h^{\mu\nu}\chi_\nu\,,\qquad \delta_\chi h_{\mu\nu}=\tau_\mu \chi_\nu+\tau_\nu\chi_\mu-2\tau_\mu\tau_\nu\tau^\rho\chi_\rho\label{htgaugesym}
\end{equation}   
To avoid introducing new degrees of freedom we are thus forced to consider the above transformation as a gauge symmetry.  The introduction of the new fields $h_{\mu\nu}$ and $\tau^\mu$ is very convenient however as they allow to solve the second constraint in \eqref{mformconstraints} explicitly:
\begin{equation}
\gnc^{\lambda}_{\mu\nu}=\tau^\lambda\partial_{\mu}\tau_\nu+\frac{1}{2}h^{\l\r}(\partial_{\m}h_{\n\rho}+\partial_{\n}h_{\m\rho}-\partial_{\r}h_{\m\n})-h^{\l\r}K_{\r(\mu}\tau_{\nu)}\label{nccon}
\end{equation}
This solution is not uniquely determined in terms of $h^{\mu\nu}$ and $\tau_\mu$ as it includes an arbitrary two-form $K_{\mu\nu}=-K_{\nu\mu}$. This two-form contains those degrees of freedom of $\gnc$ that are indepenent of $h^{\mu\nu}$ and $\tau_\mu$. As the definition of $K_{\m\nu}$ depends on a choice of $h_{\mu\nu}$ and $\tau^\mu$ it changes under the symmetry \eqref{htgaugesym}:
\begin{equation}
\delta_\chi\gnc_{\m\n}^\lambda=0\qquad \delta_\chi K_{\mu\nu}=-\partial_{\mu}\chi_\nu+\partial_{\nu}\chi_\mu+\tau_\mu\partial_\nu(\tau^\rho\chi_\rho)-\tau_\nu\partial_\mu(\tau^\rho\chi_\rho)
\end{equation}
Note that although $K_{\mu\nu}$ is part of a connection it transforms as a tensor under coordinate transformations:
\begin{equation}
\delta_\xi K_{\mu\nu}=\pounds_\xi K_{\mu\nu}
\end{equation}

\paragraph{Gauge-fixing}
Restricting ourselves to a local patch for simplicity the constraint $\partial_{[\mu}\tau_{\nu]}=0$ implies we can write $\tau_\mu=\partial_\mu t(x)$. Using the $(1,d)$ dimensional diffeomorphism invariance we can then choose $\tau_\mu=\delta_{\mu}^0$, i.e $t=x^0$, and we accordingly split the index $\mu=(0,i)$. By the constraints this gauge choice puts
\begin{equation}
h^{0\mu}=0\qquad \tau^0=1\qquad h^{ij}h_{jk}=\delta_k^i \qquad \tau^i=-h^{ij}h_{j0}\qquad h_{00}=h_{0i}h_{0j}h^{ij}
\end{equation} Furthermore this gauge-fixes the $(1,d)$ dimensional diffeomorphisms to time dependent $d$ dimensional diffeomorphisms (and constant time shifts): $\xi^\mu=(\xi^0,\xi^i(t,x))$.

We can simplify further by also using the $\chi$ transformation \eqref{htgaugesym}. Since $\delta_\chi h_{0i}=\chi_i$ we can always put $h_{0i}=0$. The constraints then furthermore imply that $\tau^i=0$ and $h_{i0}=0$. This second choice of gauge is preserved by the combination of diffeomorphisms and $\chi$ transformations that satisfy the relation
\begin{equation}
\chi_i=-h_{ij}\dot{\xi}^j
\end{equation}
The remaining fields, where for convenience we define  $G_i=K_{i0}$, then transform under these transformations as
\begin{eqnarray}
\delta h_{ij}&=&\call_\xi h_{ij}\\ \delta G_{i}&=&\call_{\xi}G_{i}+(K_{ij}-\dot h_{ij})\dot{\xi}^j-h_{ij}\ddot\xi^j \label{Ggauge}\\ \delta K_{ij}&=&\call_{\xi}K_{ij}+\partial_i(h_{jk}\dot\xi^k)-\partial_j(h_{ik}\dot\xi^k)\label{Kgauge}
\end{eqnarray}

One can now work out the equations of motion \eqref{ncmetriceom} in this gauge. As a first step one computes that the connection decomposes as
\begin{equation}
\gnc^{0}_{\mu\nu}=0\qquad \gnc^{i}_{00}=-G^i\qquad
\gnc^{i}_{0j}=\frac{1}{2}h^{ik}(\dot h_{kj}-K_{kj})\qquad
\gnc_{ij}^{k}=\Gamma_{ij}^k
\end{equation}

Some additional computation then reveals that the Trautman condition, i.e the second of these equations, is equivalent to
\begin{equation}
\nabla_{[0}K_{ij]}=0\qquad\qquad \nabla_{[i}K_{jk]}=0
\end{equation}
This can be explicitly solved by introducing two new fields: $\Phi$ and $C_i$ and writing
\begin{equation}
G_i=K_{i0}=-\nabla_i\Phi-\dot C_i\qquad K_{ij}=\partial_iC_j-\partial_jC_i
\end{equation}
Note that these fields are only defined up to a new gauge transformation:
\begin{equation}
\delta_\lambda \Phi=-\dot \lambda\qquad\qquad \delta_\lambda C_i=\partial_i \lambda
\end{equation}
The other transformations (\ref{Ggauge}-\ref{Kgauge}) furthermore imply that $\Phi$ and $C_i$ transform exactly as in (\ref{transfoPhi}-\ref{transfoC}).

Finally a straightforward although somewhat cumbersome application of algebra shows that indeed the first equation of \eqref{ncmetriceom} is equivalent to (\ref{eomh}-\ref{eomPhi}).

\subsection{Frame formulation of Newton-Cartan gravity}
In this subsection we mainly follow \cite{Andringa:2010it} to which we refer for further details and references.
\paragraph{Setup}
In this case one starts with the fields $\tau_\mu, e_\m^a, \omega_\mu{}^{ab}, \varpi_\mu{}^a, C_\mu$, they have the following local gauge transformations:
\begin{eqnarray}
  \delta \tau_\mu &=&  \partial_\mu \xi \label{taugauge}\\
  \delta e_\mu^a &=& \partial_\mu \xi^a + \Lambda^a{}_b e_\mu^b - \xi^b \omega_\mu{}^a{}_b - \xi \varpi_\mu{}^a + \chi^a \tau_\mu \\
  \delta \omega_\mu{}^{ab} &=& \partial_\mu \Lambda^{ab} + 2 \varpi_\mu{}^{[a}{}_c\Lambda^{b]c} \\
  \delta \varpi_\mu{}^a &=& \partial_\mu \chi^a + \Lambda^a{}_b\varpi_\mu{}^b - \chi^b\omega_\mu{}^a{}_b \\
  \delta C_\mu &=& \partial_\mu \lambda -\chi_b e_\mu^b + \xi_b \varpi_\mu{}^b 
\end{eqnarray}
These transformations generate the so called Bargmann algebra \cite{Duval:1984cj}, a central extension of the Gallilei algebra. For each field one can construct a gauge-covariant curvature in the standard way:
\begin{eqnarray}
R^{(\tau)}{}_{\mu \nu}&=& \partial_{\mu} \tau_{\nu}-\partial_{\nu} \tau_{\mu}\\
R^{(e)}{}_{\mu \nu}{}^a &=& \partial_\mu e_{\nu}^a -\omega_\mu{}^{a}_{b} e_\nu^b + e_\mu^b \omega_\nu{}^a{}_b + \tau_\mu \varpi_\nu{}^a - \varpi_\mu{}^a \tau_\nu\\
R^{(\omega)}{}_{\mu \nu}{}^{ab}&=& \partial_\mu \omega_\nu{}^{ab} - \partial_\nu \omega_\mu{}^{ab}- 2 \omega_\mu{}^{c[a} \omega_\nu{}^{b]}{}_c\\
R^{(\varpi)}{}_{\mu \nu}{}^a&=& \partial_\mu \varpi_\nu{}^a - \partial_\nu \varpi_\mu{}^a - \omega_\mu{}^a{}_b\varpi_\nu{}^b + \varpi_\mu{}^b\omega_\nu{}^a{}_b \\
R^{(C)}{}_{\mu \nu} &=& \partial_\mu C_\nu - \partial_\nu C_\mu + \varpi_\mu{}_b e_\nu^b - e_\mu^b \varpi_\nu{}_b
\end{eqnarray}
As a first step one reduces the number of independent fields by imposing three curvature constraints:
\begin{equation}
R^{(\tau)}{}_{\mu \nu}=0\qquad R^{(e)}{}_{\mu \nu}{}^a=0\qquad R^{(C)}{}_{\mu \nu}=0\label{curvconstr}
\end{equation}
The two last ones allow to solve for $\omega$ and $\varpi$ in terms of the other fields:
\begin{eqnarray}
  \omega_\mu{}^a{}_b &=& \partial_{[\mu} e_{\nu]}^a e^\nu_b +  \partial_{[\rho} e_{\mu] b} e^{\rho a} -  \partial_{[\nu} e_{\rho]}^c e_{\mu c}e^{\rho a}e^\nu_b + \tau_\mu e^{\rho a}e^{\nu}_b \partial_{[\rho}C_{\nu]}  \label{omegasol}\\
\varpi_\mu{}^a &=& \tau_\mu e^{\nu a} \tau^\rho \partial_{[\nu} C_{\rho]} + e^{\nu a}\partial_{[\nu} C_{\mu]} + e^{\rho a} e_{\mu b} \tau^\nu \partial_{[\rho} e_{\nu]}^b + \tau^\rho \partial_{[\mu}e_{\rho]}^a 
\end{eqnarray}
Note that this solution makes use of two additional fields $e^\mu_a$ and $\tau^\mu$. These carry however no new independent degrees of freedom as they are uniquely defined through the relations
\begin{equation}
e_\mu^{a}e^{\mu}_b=\delta^a_b\,,\quad \tau_\mu\tau^\mu=1\,,\quad e_{\mu}^a\tau^\mu=0\,,\quad \tau_\mu e^{\mu}_a=0 
\end{equation}

Note that the expression \eqref{omegasol} also defines a connection $\gnc_{\mu\nu}^\rho$ on tangent space by demanding the frame to be parallel with respect to a mixed covariant derivative:
\begin{equation}
\nabla_\mu e_{\nu}^a=\partial_\mu e_\nu^a- \gnc^\rho_{\mu \nu}e^a_\rho - \omega_\mu{}^a{}_b e_\nu{}^b - \varpi_\mu{}^a \tau_\nu =0
\end{equation}
This connection is exactly the same as \eqref{nccon} under the identifications
\begin{equation}
h^{\mu\nu}=e^\mu_ae^\nu_b\delta^{ab}\,,\qquad h_{\mu\nu}=e_\mu^ae_\nu^b\delta_{ab}\qquad K_{\mu\nu}=\partial_\mu C_\nu-\partial_\nu C_\mu \label{link}
\end{equation}

Finally, in this formulation the dynamics is given by the following equations:
\begin{equation}
e^\nu_aR^{(\varpi)}{}_{\mu\nu}{}^a=4\pi \GN\rho\,\tau_\mu\qquad e^\mu_ae^\nu_bR^{(\omega)}{}_{\mu\nu}{}^{ab}=0\label{framedyn}
\end{equation}

\paragraph{Gauge fixing}
The first of the constraints \eqref{curvconstr} is $\partial_{[\mu}\tau_{\nu]}=0$ so we can locally always choose $\tau_\mu=\partial_\mu t$. We can then use the gauge transformation \eqref{taugauge} to put $\tau_\mu$ into the form $\tau_\mu=\delta_{\mu}^0$. Note that this implies $e^{0}_a=0$. This choice of gauge has fixed the $\xi$ gauge transformations up to $\xi=$cst. Now note that we can use the $\chi^a$ gauge transformation, corresponding to a local translation, to put the $e^a_0=0$. This implies then that $\tau^\mu=\delta^\mu_0$ and and furthermore $e_i^ae_a^j=\delta_i^j, e_i^ae_b^i=\delta_a^b$. It further fixes the $\chi^a$ transformations up to those that can be undone by a compensating combination of local translations, rotations and constant time shifts. Working in terms of spatial diffeomorphisms instead of local translations by introducing $\xi^a=e^a_i\xi^i$ the gauge choice we made is invariant under transformations satisfying
\begin{equation}
\lambda^a=-e^a_i\dot\xi^i+\omega_i{}^a\xi^i+\zeta\varpi_0{}^a
\end{equation}
Plugging this into the gauge transformations of the remaining fields, where we write $C_0=-\Phi$ one finds after some algebra that:
\begin{eqnarray}
\delta e_i^a&=&\call_\xi e_i^a+\Lambda^{a}{}_be^b_i\\
\delta \Phi&=&-\dot\lambda+\call_\xi\Phi-C_i\dot\xi^i\\
\delta C_i&=&\partial_i\lambda+h_{ij}\dot \xi^j
+\call_\xi C_i
\end{eqnarray}
These are exactly (\ref{transfoPhi}-\ref{transfoh}) via \eqref{link}.

The dynamical equations \eqref{framedyn} become in this gauge\footnote{The equation $R^{(\omega)}{}_{0a}{}^a{}_b = 0$ follows automatically from the others by a Bianchi identity on the curvatures.}
\begin{eqnarray}
  R^{(\varpi)}{}_{0a}{}^a &=& 4 \pi G \rho  \\ 
  R^{(\varpi)}{}_{ia}{}^a &=& 0 \\ 
  R^{(\omega)}{}_{ia}{}^a{}_b&=& 0 
\end{eqnarray}
These are seen to be equivalent to (\ref{eomh}-\ref{eomPhi}) after working through some algebra.

\subsection{Galilean electromagnetism}
Just as in the relativistic case the theory can be expressed in terms of a one-form potential $A_\mu$ that transforms as
\begin{equation}
\delta A_\mu=\pounds_\xi A_\mu+\partial_\mu \zeta \label{vfield}
\end{equation}
The gauge invariant fieldstrength is as usual $F_{\m\n}=\partial_\mu A_\nu-\partial_\nu A_\mu$.
The equations of motion on an arbitrary Newton-Cartan background are then
\begin{equation}
h^{\mu\nu}\nabla^{(\mathrm{nc})}_\mu F_{\nu\rho}=j_{(\mathrm{e})\rho}\qquad \partial_{[\mu}F_{\nu\rho]}=j^{(\mathrm{m})}_{\mu\nu\rho}
\end{equation}
It takes only a short calculation to show that by the gauge fixing of appendix \ref{mform} the above equations and transformation rules coincide with (\ref{max1},\ref{max2}) and (\ref{transfopsi},\ref{transfoA}), when one makes the identifications $A_0=-\Psi$, $\rho_{(\mathrm{e})}=j_{(\mathrm{e})0}$ and $\rho^{(\mathrm{m})}_{ij}=j^{(\mathrm{m})}_{ij0}$. 

\subsection{Newton-Cartan-Maxwell dilaton theory}
In section \ref{nonrelsec} we derived a family of nonrelativistic theories, parameterized by a coupling constant $q$, that generalizes  $(1,d)$-dimensional Newton-Cartan gravity to include scalar and vector fields. When $q=0$ this theory is the Kaluza-Klein reduction of $(1,d+1)$-dimensional Newton-Cartan gravity. The theory contains in addition to the Newton-Cartan fields \eqref{ncfields} a vector field  \eqref{vfield} and a scalar field: $\delta \Omega=\pounds_\xi \Omega$. Its manifest $(1,d)$ diffeomorphism invariant equations of motion are given by
\begin{eqnarray}
h^{\mu\nu}\nabla_\mu^{\nc}\partial_\nu\Omega &=&\frac{1}{4}\Omega^{q+3}h^{\mu\nu}h^{\rho\sigma}F_{\mu\r}F_{\n\sigma}\\
h^{\mu\nu}\nabla_\mu^{\nc}\left(\Omega^{2q+3}F_{\nu\rho}\right)&=&0\\
\Omega R^\nc{}_{\mu\nu}&=&\nabla_\mu^\nc\partial_\nu\Omega+q\Omega^{-1}\partial_\mu\Omega\partial_\nu\Omega+\frac{1}{2}\Omega^{q+3}h^{\r\s}F_{\mu\rho}F_{\nu\sigma}\\
h^{\lambda[\mu}R^{\nc\,\nu]}{}_{(\rho\sigma)\lambda}&=&0
\end{eqnarray}
After the same gauge fixing procedure as in section \ref{mform} these equations can be shown to be identical to \eqref{nonreleqs}.

\bibliographystyle{utphys}
\bibliography{NC_lit}
\end{document}